\documentclass[aps,prd,twocolumn,nofootinbib,superscriptaddress,10pt,floatfix]{revtex4-1}
\usepackage{amstext}
\usepackage{amssymb}
\usepackage{amsmath}
\usepackage[utf8]{inputenc}
\usepackage{color}
\usepackage{graphicx}
\usepackage{array}
\usepackage{subfigure}
\usepackage{color}
\usepackage{listings}
\usepackage{natbib}
\usepackage{url}
\usepackage{bm}
\usepackage{multirow}
\usepackage{etoolbox}
\usepackage[utf8]{inputenc}
\usepackage[nolist]{acronym}
\usepackage[breaklinks, plainpages=false, colorlinks=true, anchorcolor=cyan, linkcolor=magenta, citecolor=cyan, urlcolor=purple, bookmarks=false]{hyperref}
\usepackage[normalem]{ulem}
\usepackage{tabu}
\usepackage{tikz}
\usetikzlibrary{arrows,shapes,trees,decorations.pathreplacing}
\usepackage{import}
\usepackage{svn-multi}

\newcommand{\GW}{\ensuremath{\mathrm{GW}}}

\begin{document}
\title{First gravitational-wave search for intermediate-mass black hole mergers \\ with higher order harmonics}

\author{Koustav Chandra}
    \email{koustav.chandra@iitb.ac.in}
    \affiliation{Department of Physics, Indian Institute of Technology Bombay, Powai, Mumbai 400 076, India}
\author{Juan Calder\'{o}n Bustillo}
 	\email{juan.calderon.bustillo@gmail.com}
	\affiliation{Instituto Galego de F\'{i}sica de Altas Enerx\'{i}as, Universidade de
Santiago de Compostela, 15782 Santiago de Compostela, Galicia, Spain}
	\affiliation{Department of Physics, The Chinese University of Hong Kong, Shatin, N.T., Hong Kong}
\author{Archana~Pai}
    \email{archanap@iitb.ac.in}
    \affiliation{Department of Physics, Indian Institute of Technology Bombay, Powai, Mumbai 400 076, India}
\author{I.~W.~Harry}
    \affiliation{University of Portsmouth, Portsmouth, PO1 3FX, United Kingdom}
    \email{ian.harry@ligo.org}

\keywords{intermediate-mass black hole --- gravitational waves}

\begin{abstract}
Current matched-filter searches for gravitational waves from binary black-hole mergers compare the calibrated detector data to waveform templates that omit the higher-order mode content of the signals predicted by General Relativity. However, higher-order emission modes become important for highly inclined asymmetric sources with masses above $\simeq 100 M_\odot$, causing current searches to be ill-suited at detecting them. We present a new gravitational-wave search that implements templates including higher-order modes, adapted signal-glitch discriminators, and trigger-ranking statistics to specifically target signals displaying strong higher modes, corresponding to nearly edge-on sources with total redshifted masses in the intermediate-mass black-hole range $M_T(1+z) \in (100,500) M_\odot$ and mass-ratios  $q\in(1,10)$. Our search shows a volumetric sensitivity gain of up to 450\% to these signals compared to existing searches omitting higher-order modes. We deploy our search on public data from the third observing run of Advanced LIGO. While we find no statistically significant candidates beyond those already reported elsewhere, our search sets the stage to search for higher-mode rich signals in future observing runs. The efficient detection of such signals is crucial to performing detailed tests of General Relativity, observing strong-field phenomena, and maximizing the chances of observing the yet uncharted realm of intermediate-mass black hole binaries.
\end{abstract}

\maketitle

\acrodef{BBH}[BBH]{binary black hole}
\acrodef{BH}[BH]{black hole}
\acrodef{LVC}[LVC]{LIGO Scientific and Virgo Collaborations}
\acrodef{GW}[GW]{gravitational wave}
\acrodef{CI}[CI]{confidence interval}
\acrodef{IMBH}[IMBH]{intermediate-mass black hole}
\acrodef{SNR}[SNR]{signal-to-noise ratio}
\acrodef{FAR}[FAR]{false alarm rate}
\acrodef{PSD}[PSD]{power spectral density}
\acrodefplural{PSD}[PSDs]{power spectral densities}
\acrodef{LVK}[LVK]{LIGO Scientific, Virgo and KAGRA}
\acrodef{GR}[GR]{General Relativity}
\acrodef{FF}[FF]{fitting factor}
\acrodef{O3}[O3]{third observing run}
\acrodef{IFAR}[IFAR]{inverse false alarm rate}
\acrodef{BHB}[BHB]{black hole binary}
\acrodefplural{BHB}[BHBs]{black hole binaries}

\section{Introduction} 
\label{sec:intro}

The population of binary black hole mergers identified by the current generation ground-based gravitational-wave detector network has helped us uncover several interesting and unexpected features about the intrinsic properties of \aclp{BH} in the local Universe~\citep{LIGOScientific:2018mvr, LIGOScientific:2020ibl, LIGOScientific:2021djp}. While we expect that most of these merger components are stellar remnants, there are certain \acl{GW} events whose origin can also be explained by a hierarchical formation pathway~\citep{Rodriguez:2020viw, Zevin:2020gbd, Fragione:2020han, Kimball:2020qyd, Baibhav:2021qzw,  Gerosa:2021mno, Tagawa:2021ofj}. In this formation channel, second (or higher) generation \aclp{BH} merge to form increasingly massive \acl{BH} depending on the properties of the host environment. As a result, hierarchically assembled \aclp{BH} can populate the pair-instability mass gap and can help explain the formation of \aclp{IMBH} in the local Universe~\citep{Fishbach:2017dwv, Gerosa:2017kvu, Rodriguez:2019huv, Kimball:2020opk, Doctor:2021qfn, Mapelli:2021syv, Fragione:2021nhb}. 

\begin{figure*}[htb]
    \begin{center}
        \includegraphics[width=0.99\textwidth]{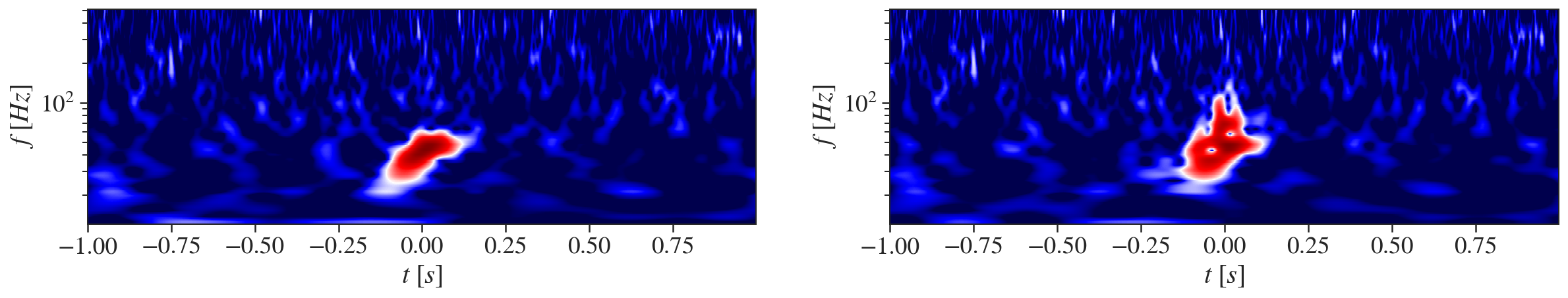}
        \caption{Time-frequency maps of two different non-precessing \acl{BBH} signals with weak higher-order modes (left) and strong higher modes (right) injected into Advanced LIGO Livingston data. The left panel corresponds to a signal from a face-on \acl{BBH} $(\iota = 0)$ with a mass ratio of $q \sim 1.2$ while the right panel corresponds to an edge-on $(\iota = \pi/2)$  binary with a mass ratio of $ q \sim 10.1$. In both cases, we have used a detector frame total mass of $M_T(1+z) \sim 500 M_\odot$. The left panel displays a (short) chirp morphology, while the right shows a more complex (multi-chirp) structure due to its higher-mode content. The lack of inspiral in the above panels is due to the high total mass of the system.}
        \label{fig:cwt}
    \end{center}
\end{figure*}

Binaries producing \acl{IMBH} remnants are particularly interesting. Firstly, the merger remnants could be seeds that grow into supermassive \aclp{BH} in the galactic nuclei, thus providing essential feedback on galaxy evolution~\citep{Quinlan:1987qj, Greene:2020}. Secondly, because of their large total mass ($\sim \mathcal{O}(10^2~M_\odot$)), the frequency of the \acl{GW} produced during the merger, and the ringdown phase is near the most sensitive band of existing \acl{GW} detectors, therefore offering the best scenarios to study the behavior of gravity in its most extreme regime~\citep{LIGOScientific:2016lio, Carullo:2019flw, Isi:2019aib}. Furthermore, during these stages, subdominant \acl{GW} emission modes can get strongly triggered depending on the properties of the source. If observed, these modes can provide crucial information on the behavior of the final object, enabling to test the nature of it through \acl{BH} spectroscopy~\citep{Dreyer:2003bv, Berti:2016lat, Bhagwat:2017tkm, Carullo:2019flw, Isi:2019aib, Cabero:2019zyt, Bustillo:2020buq} or the observation of phenomena of crucial astrophysical relevance like gravitational recoil~\citep{CalderonBustillo:2018zuq, CalderonBustillo:2019wwe, Varma:2020nbm, CalderonBustillo:2022ldv}.

The full inspiral-merger-ringdown \acl{GW} emission of compact binary mergers can be computed through either semi-analytical \citep{Santamaria:2010yb, Khan:2015jqa, Husa:2015iqa, Buonanno:2000ef, Damour:2012ky, Buonanno:1998gg, Cotesta:2018fcv} or numerical techniques \cite{Szilagyi:2009qz, Jani:2016wkt, 2017CQGra..34v4001H}. Therefore, such \aclp{GW} can be extracted from the noisy detector data through the optimal method of matched-filtering~\citep{Wainstein:1962vrq, Sathyaprakash:1991mt, Allen:2005fk}. This is the cross-correlation of the detector data with pre-computed waveform templates. For the effectiveness of matched-filtering, the search templates need to be faithful representations of the incoming \acl{GW} signal; otherwise, the searches might miss them.

However, while General Relativity predicts that gravitational waves are a superposition of several emission modes $h_{\ell,m}$, current template-based searches only implement the dominant quadrupole modes, given by $h_{2,\pm 2}$ ~\citep{Usman:2015kfa, Venumadhav:2019tad, Sachdev:2019vvd, Aubin:2020goo, Chu:2020pjv}. Such a strategy has been demonstrated to effectively detect signals from face-on (or face-off) systems with nearly equal mass and total redshifted masses $\lesssim 100M_\odot$, for which non-quadrupolar (or higher-order) modes contribute negligibly. Also, such \textit{optimally-oriented}, symmetric sources are intrinsically luminous and hence easier to detect. However, asymmetric black-hole binaries emit gravitational waves with strong higher harmonics, especially during the merger and ringdown stages~\citep{Mills:2020thr} (See Figure~\ref{fig:cwt} for a qualitative example). The contribution of higher modes in the observed signal increases as the orbital inclination of the system deviates from face-on/off, especially impacting those of high mass, for which the frequency of the dominant harmonic lies below the optimal sensitivity of the detector. Their omission in searches, therefore, dramatically reduces the sensitivity to these sources ~\citep{Capano:2013raa, CalderonBustillo:2015lrt, CalderonBustillo:2016rlt,  CalderonBustillo:2017skv, Chandra:2020ccy}, potentially causing an ``observational bias'' against asymmetric \acl{IMBH} binaries with large orbital inclinations.

~\citet{Harry:2017weg} developed a prototype matched-filter-based search for non-spinning \acl{BBH} sources containing higher-order harmonics, demonstrating its ability to recover synthetic signals in simulated Advanced LIGO noise that is free of instrumental transients or ``glitches''. These noisy artifacts tend to affect search efficiency severely. Here, we turn this prototype search into a fully working one and deploy it on data collected during the third observing run of Advanced LIGO. {\it First}, we expand the search to the case of aligned-spin (non-precessing) sources. {\it Second}, to make the search effective when applied to real data, we adapt existing signal-glitch discriminators to separate the noisy transients from astrophysical ones better. While we do not find any new statistically significant signal candidates beyond those already reported elsewhere~\citep{LIGOScientific:2020ibl, LIGOScientific:2021djp, Chandra:2021wbw, Nitz:2021uxj, Nitz:2021zwj, Olsen:2022pin},  we show that our search is up to 450\% more sensitive than past searches with overlapping parameter space, mainly when the target sources are at nearly edge-on orientation.

The rest of this paper is organized as follows. In Sec.~\ref{sec:background} we review the fully generic search method. Sec.~\ref{sec:methods} presents the details of our coincidence analysis with two detectors, and Sec.~\ref{sec:sensitivity} assesses the benefits of deploying our search. In Sec.~\ref{sec:results} we present the results of our search on O3 data, and we conclude in Sec.~\ref{sec:conclusions}.

\section{Background}
\label{sec:background}

\begin{figure}[htb]
    \begin{center}
        \includegraphics[width=\columnwidth]{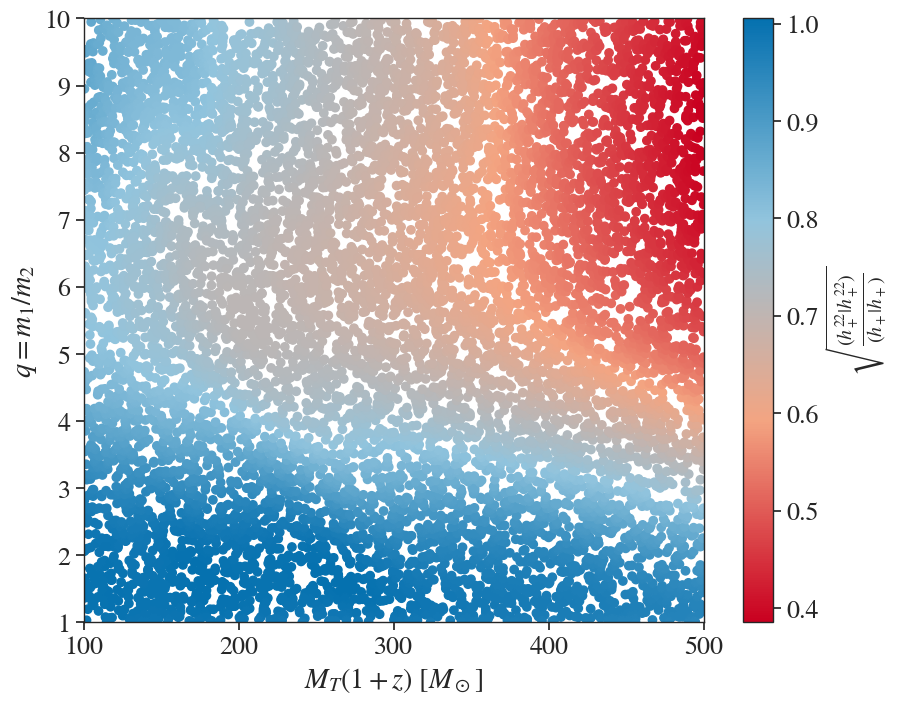}
        \includegraphics[width=\columnwidth]{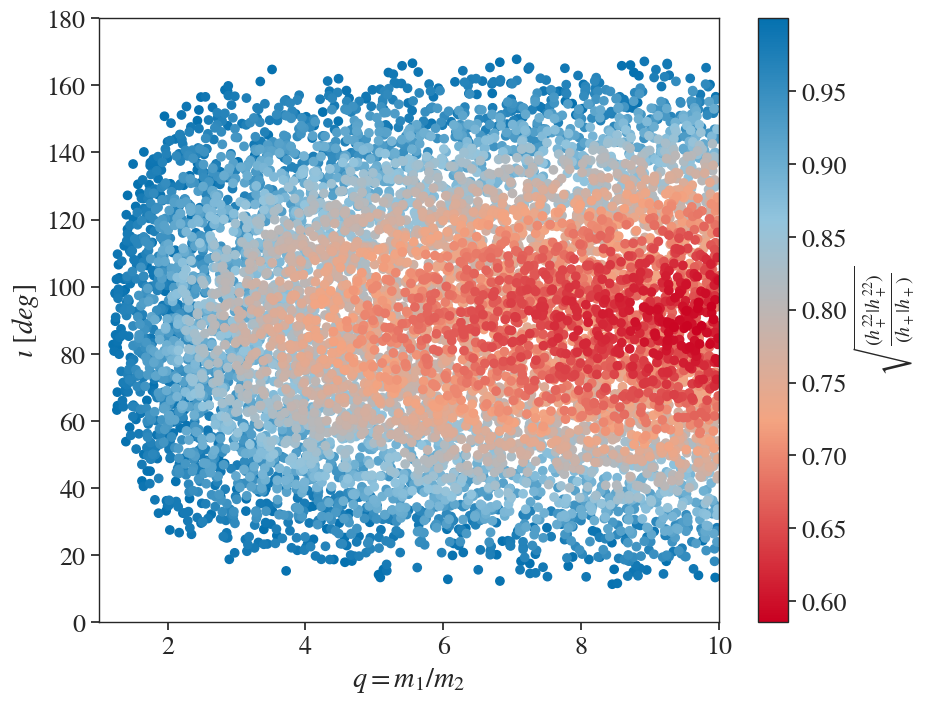}
        \caption{Ratio of the optimal \acl{SNR} of waveforms including only the dominant $(\ell,m)=(2,\pm 2)$ mode and including the modes $(\ell,|m|)=(2,1),~(2,2),~(3,3),~(4,4)$ and $(5,5)$, generated with identical parameters. The top panel shows results for varying the source's total mass and mass ratio for fixed edge-on inclinations. The bottom panel shows results for varying mass ratio and inclination for a fixed redshifted total mass of $M_T(1+z)=300~M_\odot$. In both panels, we have set the black hole spins to zero, the azimuth to $\phi=0$, and the antenna response $F_+=1$ and $F_\times=0$. We consider a detector characterized by the Advanced LIGO Livingston sensitivity during the \ac{O3} run.}
        \label{fig:ratio}
    \end{center}
\end{figure}

The fundamental assumption behind most \acl{GW} data analysis is that the detector output $d(t)$ is composed of two additive components, namely the noise $n(t)$ and the signal strain $s(t)$ as:
\begin{equation}
    d(t) = n(t) + s(t)
\end{equation}
The noise $n(t)$, a stochastic process, varies randomly with time. Assuming that the noise is wide-sense stationary and Gaussian, its statistical properties are fully described by its one-sided power spectral density $S_n(f)$. The \acl{GW} signal, on the other hand, is deterministic, and it will impart a strain $s(t)$ that is parameterized by a vector $\boldsymbol{\lambda}$ and is given by the following linear combination:
\begin{equation}
    s(t; \boldsymbol{\lambda}) = F_+ h_+(t-t_c; D_L, \boldsymbol{\theta}) + F_\times h_\times(t-t_c; D_L, \boldsymbol{\theta})~.
\end{equation}
Here, $F_{+/\times}$ are the sky-location and polarisation angle-dependent antenna response patterns of the detector to the two \acl{GW} polarisations, $h_{+/\times}$. $t_c$ is the merger time of the signal, and $D_L$ is the luminosity distance to the source. The morphology and the evolution of these two polarisation states depend on the properties and orientation of the source, which we denote using the vector $\boldsymbol{\theta}$. 

We can measure the loudness of a signal as it appears in a detector by calculating the optimal \acl{SNR}:
\begin{equation}
    \rho_\mathrm{opt} = \sqrt{(s|s)}
\end{equation}
Here, $(a|b)$ is the real part of the noise-weighted inner product:
\begin{equation}
    \langle a | b \rangle = 4 \int^{f_\mathrm{max}}_{f_\mathrm{min}} df~\frac{\tilde{a}^\ast(f)\tilde{b}(f)}{S_n(f)},
\end{equation}
between two real-valued time series. The tilde here denotes the Fourier transform of the corresponding time-domain data. $f_\mathrm{min}$ and $f_\mathrm{max}$ represent upper and lower frequency cutoffs, and the symbol $\ast$ denotes complex conjugation. 

In principle, we can model this signal strain for any given \acl{GW} source. This way, we can construct waveform templates $h(t;\boldsymbol{\lambda})$ for the expected signal. When templates are available, the optimal way to retrieve the signals from the noise is via matched-filtering whose output is the matched-filter \ac{SNR}~\citep{Wainstein:1962vrq, Sathyaprakash:1991mt, Allen:2005fk}:
\begin{equation}\label{eq:SNR-I}
    \rho = \frac{(d|h(\boldsymbol{\lambda}))}{(h(\boldsymbol{\lambda})|h(\boldsymbol{\lambda}))^{1/2}}~.
\end{equation}
However, the parameters $\boldsymbol{\lambda}$ are not known a priori. Therefore matched-filter-based searches construct and use discrete banks of template waveforms spanning the search space to compare the data and numerically maximize the \ac{SNR}. In certain situations, we can make simplifying assumptions about the morphology of the incoming signal and can hence analytically maximize over some of these parameters. This greatly reduces the computational cost of the search, for it reduces the dimensionality of the parameter space over which we need to maximize the \ac{SNR} numerically. 

\subsection{Gravitational Wave Higher Harmonics}

The \acl{GW} emission from a compact merger can be expressed as a superposition of different GW emission modes $h_{\ell,m}$ weighted by spin-2 spherical harmonics $^{-2}Y_{\ell,m}$~\citep{Goldberg:1966uu, Blanchet:2013haa}:
\begin{equation}
    h_+ - ih_\times = \frac{1}{D_L} \sum_{\ell =2}^\infty\sum_{m=-l}^{l}~^{-2}Y_{\ell,m}(\iota, \phi)h_{\ell,m}(t-t_c;\boldsymbol{\Xi})    
\end{equation}
Above,  $(\iota, \phi)$ define the polar and azimuthal angles of a spherical coordinate system centered at the center of mass of the system, and $\boldsymbol{\Xi}$ collectively denotes the intrinsic parameters of the source, namely the individual masses $m_{1,2}$ and spins $\vec{\chi}_{1,2}$. 

In order to isolate the contribution from each orientation parameter, is it useful to decompose the harmonics $Y^{-2}_{\ell,m}$  in terms of an overall amplitude term that depends on the source inclination and an overall phase term that depends on the azimuth as:
$$
       ^{-2}Y_{\ell,m} = A_{\ell,m}(\iota)e^{-im\phi}. 
$$
This makes it obvious that the inclination angle $\iota$ determines the amplitude of each mode while the azimuth $\phi$ determines the way the modes combine, which can dramatically change the signal morphology observed by different observers around the source. Similarly, we can also express the $(\ell,m)$ emission mode of a \acl{GW} signal as 
$$h_{\ell,m} = \mathcal{A}_{\ell,m}(t-t_c; \boldsymbol{\Xi})e^{-i\Phi_{\ell,m}(t-t_c;\boldsymbol{\Xi})}$$.

During most of the inspiral part of non-precessing quasi-circular mergers, the emission is vastly dominated by the quadrupolar mode or $(\ell, |m|)=(2, 2)$. However, the impact of higher-order modes in the signal becomes significant during the merger and ringdown stages. This effect becomes enhanced for high-mass systems, for which the frequency of the dominant harmonic can lie below the detector's sensitive band. In addition, the amplitude $\mathcal{A}_{\ell,m}$ of these modes relative to the $\mathcal{A}_{2,\pm 2}$ grows with increasing mass ratio, making the higher-order modes very relevant for asymmetric mass sources~\citep{Berti:2007fi, Mills:2020thr}. Finally,  while the spherical harmonics other than $Y^{-2}_{2,\pm 2}$ is mostly zero for face-on (or face-off) orientations $(\iota=0,\pi)$, they reach their maxima for intermediate inclinations, making the higher-order modes have a stronger impact for large orbital inclinations, e.g. for edge-on orientations $(\iota=\pi/2)$. 

To visualize the above, we, in Figure~\ref{fig:ratio}, shows the ratio of the optimal \acl{SNR} of the $h_+$ polarisations of simulated waveforms from non-spinning \acl{BBH} systems when these are generated using only the dominant quadrupole and using the modes $(\ell,|m|)=(2,1),~(2,2),~(3,3),~(4,4)$ and $(5,5)$. The top panel shows how the ratio varies as a function of the redshifted total mass and mass ratio for (fixed) edge-on inclinations. The bottom panel shows the same as a function of mass ratio and inclination for a fixed redshifted total mass of $300 M_\odot$. For approximately $51\%$ of the simulated signals, we find that the quadrupole mode contributes less than $75\%$ to the SNR, indicating that the sub-dominant harmonics significantly contribute to the overall signal loudness, especially for asymmetric, nearly edge-on sources. This makes it obvious that higher modes are crucial to represent such signals accurately.

\subsection{Generic SNR Statistic}
\label{sec:generic-statistics}

Irrespective of the parameters of the template, we can always write the template strain, in the Fourier domain, as:
\begin{equation}\label{eq:strain-I}
    \begin{aligned}
        \tilde{h}(f) &= F_+\sqrt{( h_+ | h_+)}~\hat{h}_+ + F_\times\sqrt{(h_\times | h_\times)}~\hat{h}_\times \\ 
    &=A(u\hat{h}_+ + \hat{h}_\times)
    \end{aligned}
\end{equation}
by defining:
\begin{equation}
\begin{aligned}
u= & \frac{F_+\sqrt{( h_+ | h_+)}}{F_\times\sqrt{( h_\times | h_\times)}} \\
A= & F_\times\sqrt{( h_\times|h_\times)}  \\
\hat{h}_{+/\times} = &  \frac{h_{+/\times}}{\sqrt{( h_{+/\times} | h_{+/\times})}}
\end{aligned}
\end{equation}
This indicates that it is always possible to express a generic \acl{GW} transient in terms of an overall amplitude term $A$ and a weighted linear combination of the unit-normalized \acl{GW} polarisations. 

If we substitute Eq.~\eqref{eq:strain-I} in Eq.~\eqref{eq:SNR-I}, we get:
\begin{equation}
    \rho = \frac{(d|u\hat{h}_+ + \hat{h}_\times)}{(u\hat{h}_+ + \hat{h}_\times| u\hat{h}_+ + \hat{h}_\times)^{1/2}}
\end{equation}
Maximising $\rho$ over the $u-$dependence yields~\citep{Harry:2017weg}:
\begin{equation}\label{eq:SNR-II}
    \max_{u} \rho^2 = \frac{( d | \hat{h}_+)^2 + ( d | \hat{h}_\times)^2 -2( d | \hat{h}_+)( d | \hat{h}_\times)(\hat{h}_+| \hat{h}_\times)}{1-(\hat{h}_+ | \hat{h}_\times)^2}~, 
\end{equation}
which by construction is effectively maximized over the luminosity distance to the source, the sky location of the source, and the polarisation angle. Using an inverse Fast Fourier transform routine, we can also carry out the maximization over $t_c$. Therefore, we can use this \ac{SNR} statistic to search for \textit{any \acl{GW} transient} irrespective of their morphology, provided the waveform of the transient is well-modeled. For that reason, we use this statistic to perform our search with higher-order mode templates that are completely parameterized by the vector  $\boldsymbol{\theta}=(m_1,~m_2,~\chi_{1z},~\chi_{2z},~\iota,~\phi)$.

\subsection{SNR statistic used in current searches: Non-precessing Quadrupolar Limit}
\label{sec:limit}

\begin{figure}[!ht]
    \centering
        \includegraphics[width=0.9\columnwidth]{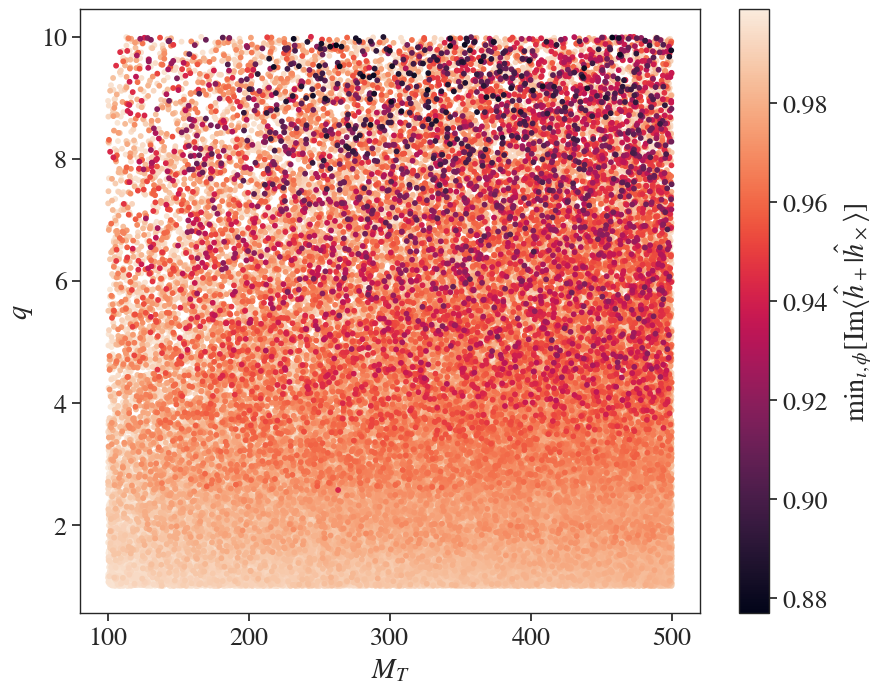}
        \includegraphics[width=0.9\columnwidth]{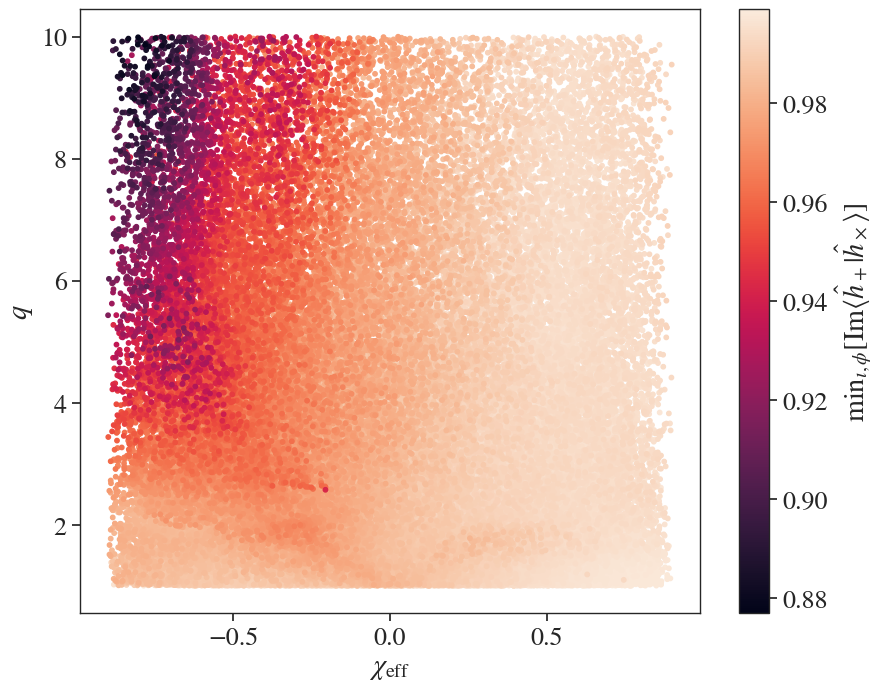}
        \includegraphics[width=0.9\columnwidth]{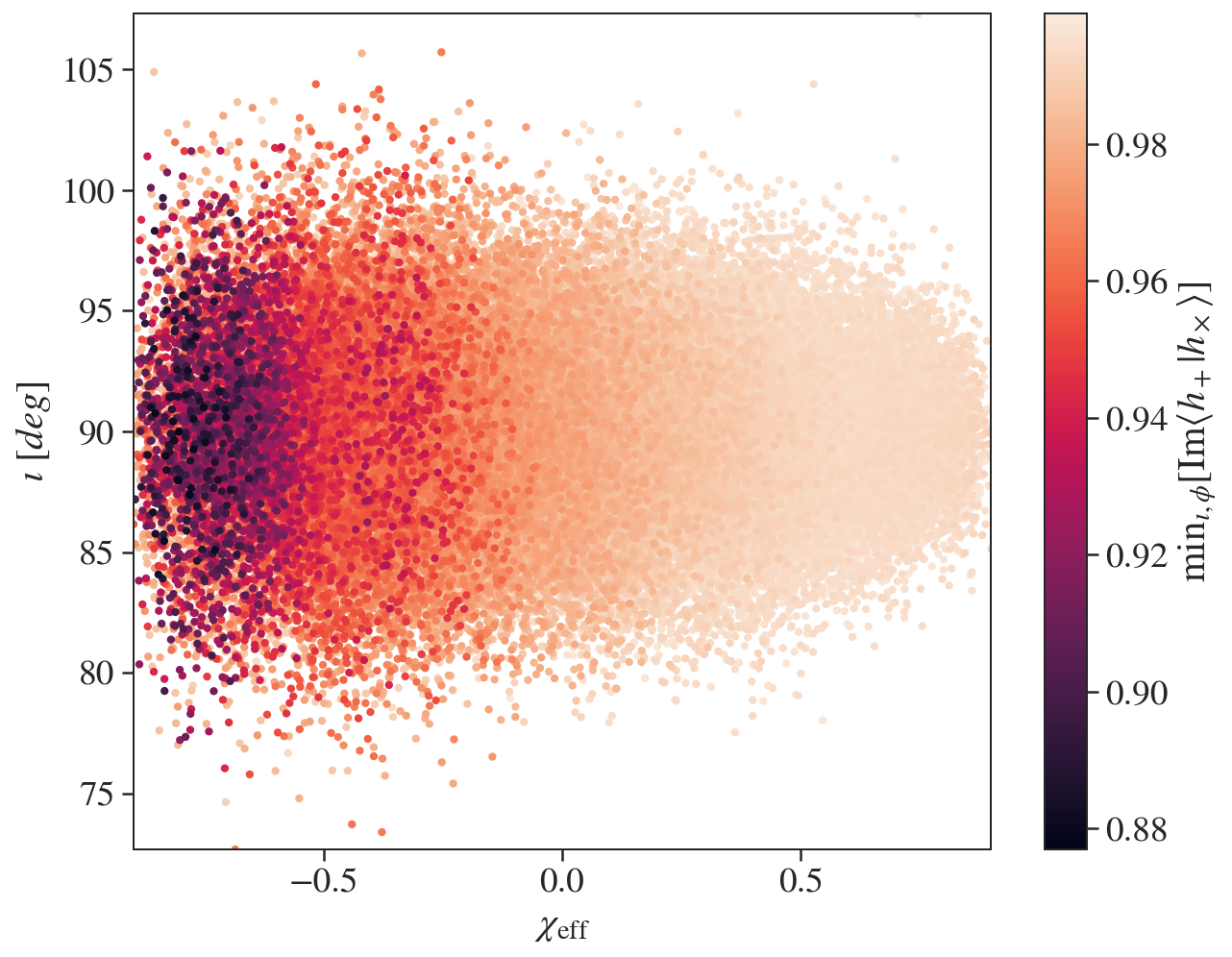}
        \caption{Magnitude of the imaginary part of the complex overlap $\langle h_+ | h_\times \rangle$ between the two polarisations for a simulated population of quasi-circular \acl{BBH} signals with higher-mode content. The top panel shows this as a function of the total mass and mass ratio, while the middle panel shows the same as a function of effective aligned spin and mass ratio. We have numerically minimized the overlap over the inclination angle and the azimuth for each system. The bottom panel shows the minimum overlap as a function of the corresponding inclination angle and effective aligned spin.}
        \label{fig:overlap}
\end{figure}

As already stated, most current matched-filter-based searches make several assumptions about the signal model, which is not true when using waveforms with higher harmonics~\citep{Usman:2015kfa, Venumadhav:2019tad, Sachdev:2019vvd, Aubin:2020goo, Chu:2020pjv}. Specifically, they use a \ac{SNR} statistic that assumes that the polarisation states are related as $\tilde{h}_\times \propto i \tilde{h}_+$. The generic \ac{SNR} statistic in Eq.~\eqref{eq:SNR-II} for such an assumption reduces to:
\begin{equation}
\label{eq:SNR-III}
    \max_{u} \rho^2 = (d|\hat{h}_+)^2 + (d|\hat{h}_\times)^2 = |\langle d | \hat{h}_{+} \rangle| ^2,
\end{equation}
as $(\hat{h}_+|\hat{h}_\times)=0$. If we limit to dominant harmonics of quasi-circular black hole binaries, the constant of proportionality is:
$$
\epsilon = - \frac{2\cos{\iota}}{1+\cos^2{\iota}}
$$
and its magnitude varies between $0$ (for edge-on) and $1$ (face-on/off).

Also, current searches assume that we can absorb the binary's sky location, orientation, luminosity distance, and corresponding polarisation angle by applying an appropriate amplitude and phase scaling to the observed waveform, both of which can be analytically maximized. Therefore, current matched-filter searches only iteratively search over the source's intrinsic parameters $\boldsymbol{\Xi}$. Both assumptions hold if and only if we restrict to the dominant harmonics of a quasi-circular \acl{BBH} merger. However, they break when we use waveforms, including higher-order modes as templates.

\subsection{Evaluating the need for the generic SNR statistic for higher-order mode searches}

We assess the necessity of the generic SNR statistic for higher-order mode searches by computing the magnitude of the imaginary component of the complex overlap between the unit-normalized gravitational-wave polarisations:
\begin{equation}
    O = \mathrm{Im}\langle \hat{h}_+ | \hat{h}_\times\rangle
\end{equation}
for O3 Advanced LIGO noise sensitivity. We perform this study with waveforms generated using the reduced-order representation of the aligned-spin effective-one-body model with higher-order modes, \texttt{SEOBNRv4HM} that includes the spherical harmonics $(\ell,|m|)=(2,1),~(3,3),~(4,4)$ and $(5,5)$ beyond the dominant quadrupolar mode~\citep{Cotesta:2018fcv, Cotesta:2020qhw}. The simulated waveforms imitate gravitational waves from a synthetic population of quasi-circular black hole binaries with detector frame (redshifted) total mass $M_T(1+z) \in (100,500) M_\odot$, mass-ratio $q=m_1/m_2 \in (1,10)$ and spins $\chi_{1z,2z}\in (-0.998, 0.998)$. We distribute these sources isotropically over the inclination angle and reference orbital phase, and then for each of these waveforms, we compute $O$. Finally, following~\citep{Harry:2017weg}, we numerically minimize $O$ over the inclination angle and phase. Fig~\ref{fig:overlap} shows the minimum value of $O$ as a function of total mass and mass ratio in the top panel and as a function of mass ratio and effective aligned spin in the middle panel.

We find that for certain configurations, the minimum value of the overlap goes to $\sim 0.87$. Also, for $\sim 3.4\%$ of the cases, $O < 0.95$, indicating that we cannot, in general, assume $\tilde{h}_+(f) \propto i\tilde{h}_\times(f)$ for quasi-circular binary waveforms. Further, most of these low overlap binaries have either a high total mass and/or a negative $\chi_\mathrm{eff}$, indicating that their duration is short within the detector bandwidth. Also, these low overlap binaries are oriented nearly edge-on, as shown in the bottom panel of Fig~\ref{fig:overlap}. Therefore, we target binaries with $\iota \in (75^\circ, 105^\circ)$ and use Eq.~\eqref{eq:SNR-II} for SNR calculation.


\section{Methods for observing generic black hole binaries}
\label{sec:methods}

There are two matched-filter analyses that explicitly search for binary black hole mergers producing lower mass range~$(100,600)~M_\odot$ intermediate-mass black hole remnants. The first of the two, namely PyCBC-IMBH~\citep{Chandra:2021wbw}, targets quasi-circular binaries with detector frame total masses between 100 and 600 $M_\odot$, with component masses greater than $40 M_\odot$ and mass ratio $q$ between $1$ and $10$. The search, however, uses waveforms with just the dominant harmonics and a matched-filter \ac{SNR} statistic defined in Eq~\eqref{eq:SNR-III}. Also, to reduce the number of false alarms due to short-duration glitches~\citep{Cabero:2019zyt}, this search does not use any template with a duration less than $70$ms, measured from the fixed starting frequency of $15$Hz.

The other intermediate-mass black hole binary-specific search is constructed using the GstLAL software package~\citep{Sachdev:2019vvd, CANNON2021100680}, and it also uses a template bank of quasi-circular dominant (2,2) harmonic with $q \in (1,10)$ and assumes that $\tilde{h}_\times \propto i \tilde{h}_+$. However, unlike PyCBC-IMBH, it targets systems with $M_T(1+z)$ between 50 and 600 $M_\odot$, and it uses a starting frequency of $10$ Hz for its matched-filter operation. Other than these differences, the searches also use different signal-noise discriminators and rank coincident triggers differently (For details, see Sec.~3.2 of ~\citet{LIGOScientific:2021tfm} and the references therein).

Both of these analyses need to incorporate knowledge of the higher harmonics into a search for gravitational waves, making them poor at detecting nearly edge-on quasi-circular binaries, which are mass asymmetric and/or massive. In what follows, we describe our search strategy that involves the construction of a higher-order mode bank and adapting existing signal-glitch discriminators to perform a search on real \acl{GW} data.

Finally, we note that search algorithms beyond those relying on matched-filtering have also been used to search for intermediate-mass black hole binaries in LIGO-Virgo data. In particular, the template-independent search algorithm Coherent WaveBurst~\citep{Klimenko:2015ypf, Szczepanczyk:2020osv} identifies \acl{GW} signals by looking for coherent power excess across different detectors using minimal assumptions on the morphology of the expected signal. This makes this search potentially sensitive toward a wider variety of intermediate-mass black hole binaries. Currently, the sensitivity of the Coherent WaveBurst search (in its intermediate-mass black hole configuration) is comparable with that of the PyCBC-IMBH search~\citep{Chandra:2021xvs}.

\subsection{Search space}
\label{sec:bank}

Motivated by Figure~\ref{fig:ratio}, we build a template bank to target highly inclined sources with redshifted total mass beyond $100~M_\odot$. The specifics of our target domain are summarised in Table~\ref{bank}.
\begin{table}[h]
    \centering
    \begin{tabular}{l r}
          Parameter &  \\
         \hline \hline
         Total redshifted mass & $M_T(1+z) \in (100,~500) M_\odot $ \\
         Mass ratio & $q = m_1/m_2 \in (1,~5)  ~\&~  (5,~10)$ \\
         Spin z-component  & $\chi_{1z,2z} \in (-0.998,~0.998)$ \\
         Orbital inclination & $\iota \in (75^\circ,~105^\circ)$ \\
         Azimuth & $\phi  \in (0,2\pi)$ \\
         \hline \hline
         \end{tabular}
         \caption{Summary of the target parameter space covered by our template banks. Note that we consider two banks ${\cal{B}}_1$ with $q \in (1,5)$ and ${\cal{B}}_2$ with $q \in (5,10)$.}
    \label{bank}
\end{table}

While this search can be expanded to larger regions of the parameter space of \aclp{BHB}, particularly with lower orbital inclinations, we have opted to restrict to the above constraints for the following reasons. Firstly, templates with lower orbital inclinations will identify intrinsically louder sources than templates with higher inclinations. Therefore, the net sensitivity of our search would be dominated by that to face-on binaries (which do not display higher modes), damaging our ability to evaluate how our search improves on existing ones when targeting higher-mode rich signals. In addition, including more templates in our bank would increase our false alarm rate due to the increased number of templates, further hindering the sensitivity to the sources we want to target. Secondly, as mentioned before, sources at a high inclination produce complex, higher-mode-rich signals, while face-on ones produce rather morphologically simple ones. Therefore, it is reasonable to expect the corresponding two types of templates to show a different propensity to be triggered by glitches.

Consequently, we preferred to isolate these two types of potential background populations. Thirdly, the upper bound on the detector frame total mass is chosen to ensure a minimum template duration, which should be larger than one cycle for effective matched filtering. In fact, in preliminary analyses, we found that templates for heavier systems are more susceptible to glitches, resulting in a poorer overall search sensitivity. Similarly, we keep the mass ratio within $q \leq 10$. Finally, due to similar arguments, we divide our search into two separate template banks: the first -- ${\cal{B}}_1$ -- targeting systems with $q \in (1,5)$ and the second  -- ${\cal{B}}_2$ -- targeting systems with $q \in (5,10)$. This helps protect the lower mass ratio space from the high penalty -- due to increased glitchiness -- of the high mass ratio region. Lastly, while less physically relevant, enlarging the target parameter space would have led to a higher number of templates, equating to higher computational costs.

\begin{figure}[htb]
    \begin{center}
        \includegraphics[width=\columnwidth]{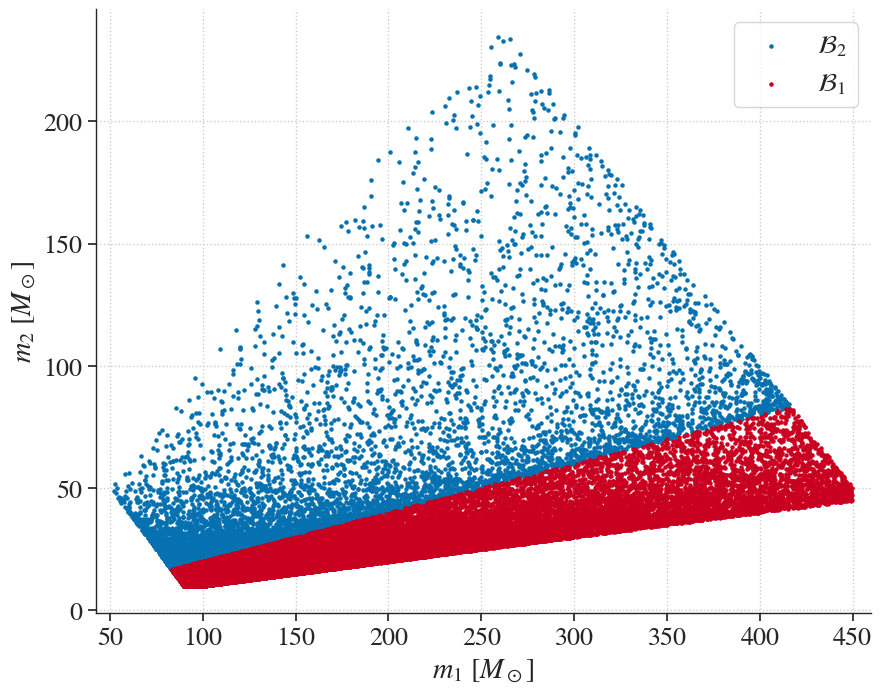}
        \includegraphics[width=\columnwidth]{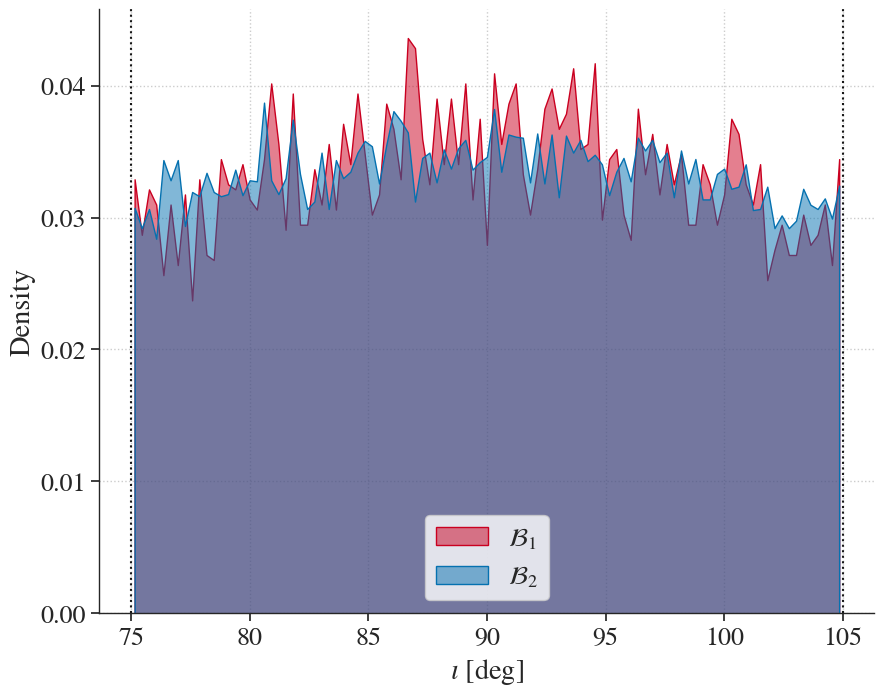}
        \caption{The top panel shows the distribution of template waveforms in the component mass space for our two banks. The red templates belong to the low-mass-ratio ($q\in[1,5]$) region (${\cal{B}}_1$), and the blue templates belong to the high mass-ratio ($q\in[5,10]$) region (${\cal{B}}_2$). The bottom panel shows the distribution of the orbital inclination angle $\iota$ of the templates, which is limited to $\iota \in (75^\circ,105^\circ)$.}
        \label{fig:bank}
    \end{center}
\end{figure}

\subsection{Template Bank}

\begin{figure}[htb]
    \begin{center}
    \includegraphics[width=\columnwidth]{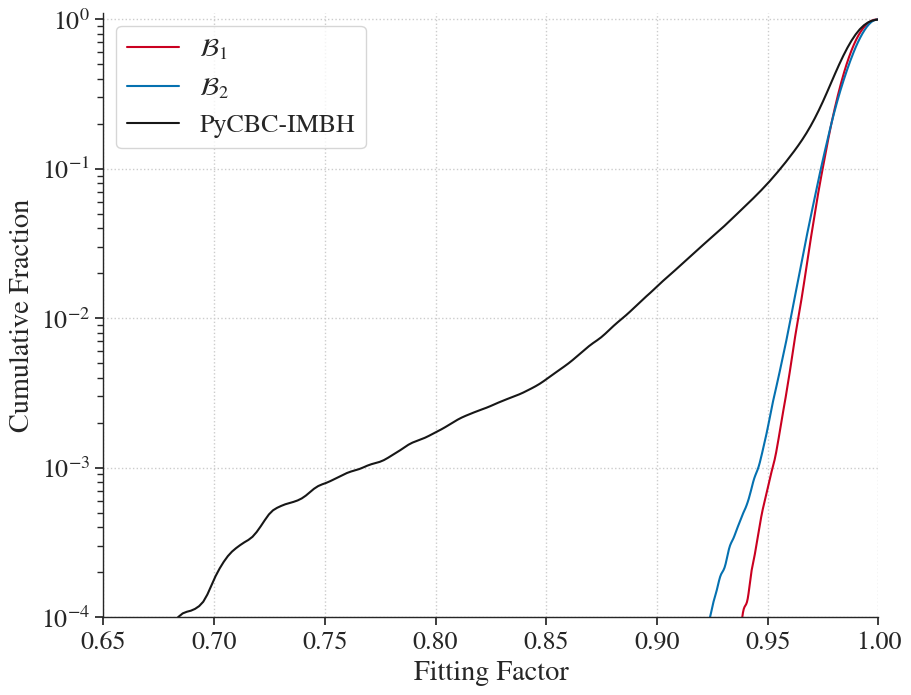}
    \caption{Effectualness of our template banks to a family of randomly generated waveforms within their respective target parameter spaces. For comparison, we also show the effectualness of the PyCBC-IMBH bank to the same set of waveforms, which spans all of our parameter space but lacks higher-order modes in its templates. As can be seen, the PyCBC-IMBH bank owing to its lack of higher-order modes, is largely un-effectual (fitting factor below 0.97) within our target space.}
    \label{fig:fitting}
    \end{center}
\end{figure}

While several such models exist, in this work, we choose the aligned-spin effective-one-body model \texttt{SEOBNRv4HM-ROM}. This model includes modes $(\ell,|m|)=(2,1),~(3,3),~(4,4)$ and $(5,5)$ beyond the quadrupolar mode \citep{Cotesta:2018fcv, Cotesta:2020qhw}. We choose a minimum frequency cutoff for the matched filter of $15$Hz. We note that, in Gaussian background noise, a lower low-frequency cutoff would increase our sensitivity to \acl{BBH} signals simply due to increased SNR due to contributions from low frequencies. In non-Gaussian noise, however, this also increases the chances that our templates are triggered by glitches, yielding a larger background that would compensate (and even overcome) the aforementioned gain in sensitivity. 

We build each of our banks, namely ${\cal{B}}_1$ and ${\cal{B}}_2$, with a stochastic template placement algorithm~\citep{Harry:2008yn, Babak:2008rb, Harry:2009ea} and a minimal match criterion of $0.97$. The template placement algorithm relies on choosing a random ``template '' $h_T$ from our target search space, checking whether the bank, $\mathfrak{B}$ has a \ac{FF}:
\begin{equation}
    \mathrm{FF} = \max_{t_c,~u,~h_i \in \mathfrak{B}} (h_T|h_i)
\end{equation}
less than $0.97$ toward the template and accepting it based on it~\citep{Apostolatos:1994mx}. If otherwise, we reject the template. We repeat this process unless we reach a sufficiently large rejection rate. The minimal match choice of $0.97$ ensures that the maximum \ac{SNR} loss due to the discreteness in the template bank is not more than $3\%$. While a denser template bank with a larger minimal match would lead to a smaller SNR loss -- e.g., the PyCBC-IMBH bank uses a minimum match of 0.99 -- and, in principle, raise the sensitivity of the search, this would also increase its computational cost, which is already high due to the increase in the number of templates coming from the addition of the orientation parameters to the bank and the need to perform two filters per template.  

Figure~\ref{fig:bank} shows the resulting template banks. The top panel shows the two banks in terms of the component masses of the \acl{BBH}. The bottom panel shows the orbital inclination distribution of the templates. While ${\cal{B}}_1$ has 8626 templates, ${\cal{B}}_2$ has 40915 templates, i.e., almost $4.7$ times larger. Consequently, ${\cal{B}}_2$ will have a larger background, making it more prone to false alarms. We discuss this impact in terms of the search sensitivity in Sec.~\ref{sec:sensitivity}. We will call the search implementing ${\cal{B}}_1$ as search-1 and the one implementing ${\cal{B}}_2$ as search-2.

We check the ``effectualness'' of our banks to their target space by calculating \acl{FF} towards $\sim 100,000$ \texttt{SEOBNRv4HM} waveforms randomly distributed across the target parameter space of the banks, and we summarise our findings in Figure~\ref{fig:fitting} where we plot our corresponding \acl{FF} distributions. We note that for both our banks, the effectualness is higher than the target value of $0.97$ for $\sim 98\%$ of the target signals. For comparison, we also show the effectualness of the bank used by the PyCBC-IMBH search towards the same set of simulated signals. The latter has a minimum recovered fitting factor of $0.64$ and falls below $0.97$ for more than $15\%$ of the target signals showing that such a bank is ineffectual in our target search space.

\subsection{Dealing with instrumental transients}
\label{sec:signal-glitch}

\begin{figure*}[htb]
    \begin{center}
        \includegraphics[width=0.49\textwidth]{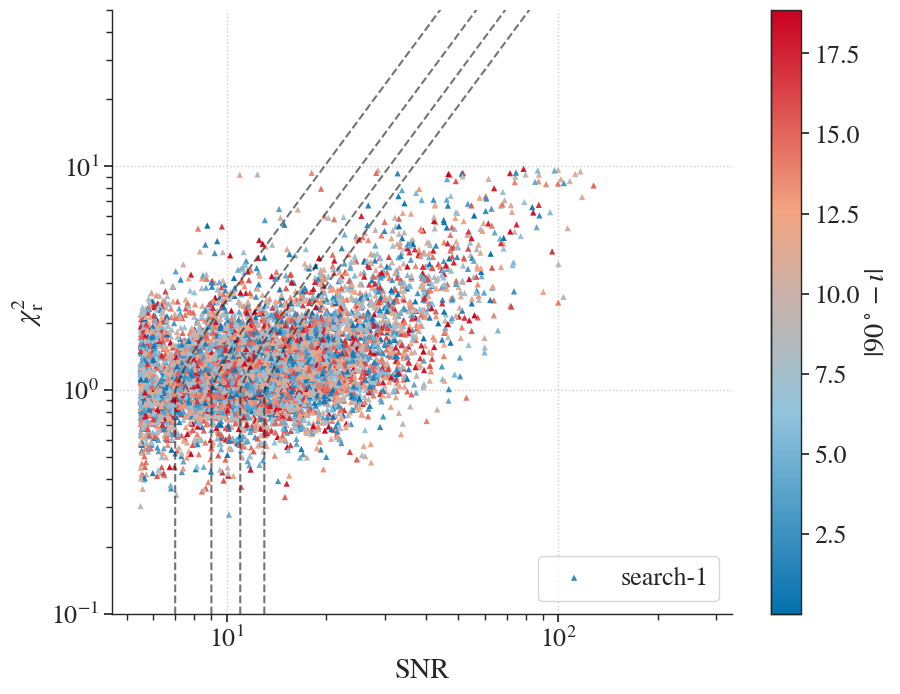}
        \includegraphics[width=0.49\textwidth]{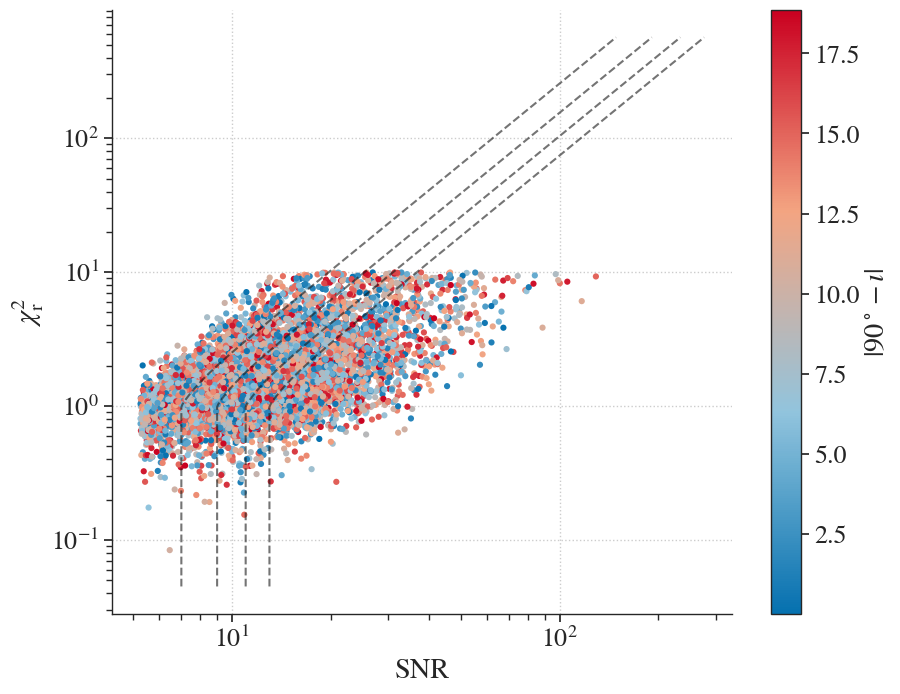} \\
        \includegraphics[width=0.49\textwidth]{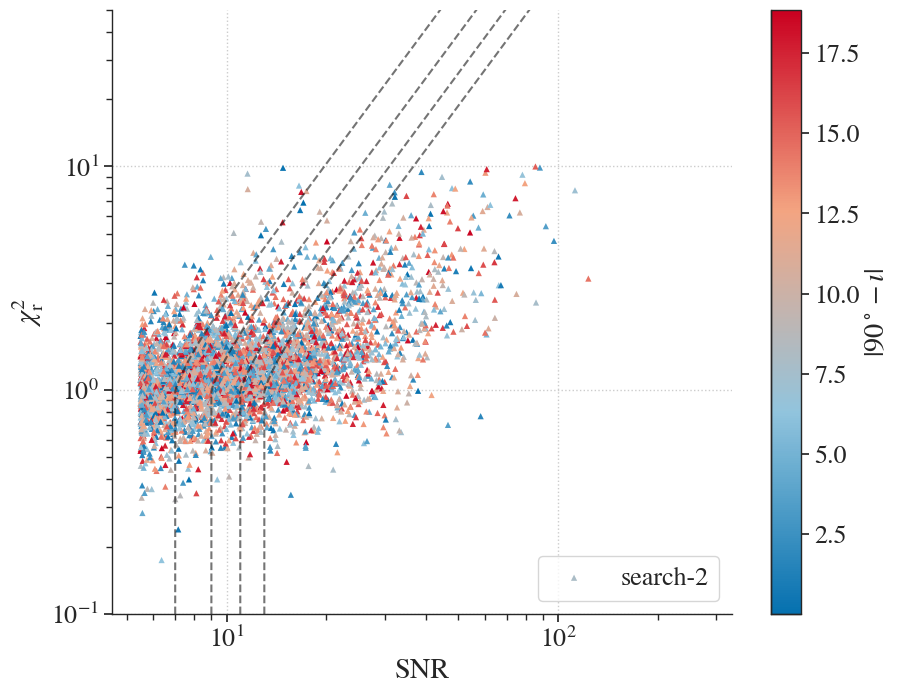}
        \includegraphics[width=0.49\textwidth]{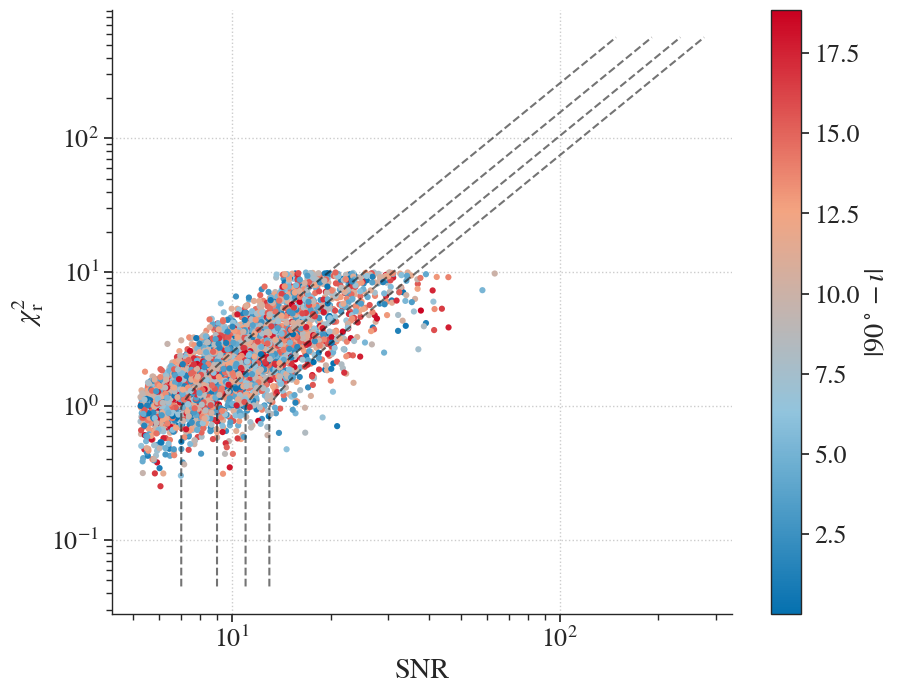}
        \caption{Impact of inclusion and omission of the higher-order modes on signal-glitch consistency tests for current searches and the one presented here. In the top panels, we plot the \ac{SNR}- $\chi^2_\mathrm{r}$ distribution for a simulated population of \acl{BBH} obeying the constraints of ${\cal{B}}_1$. In the bottom panels, we do the same for a simulated population of \acl{BBH} obeying the constraints of ${\cal{B}}_2$. The left panels show results corresponding to our banks, including higher modes, while the right panels show the results corresponding to the PyCBC-IMBH search. The dashed lines show contours of constant ranking statistic $\tilde{\rho}$. The simulated signals, which contain higher modes, have a greater mismatch with the waveforms implemented in the PyCBC-IMBH bank, leading to significantly larger $\chi^2_\mathrm{r}$-values for the same SNR. As a result, the search will misinterpret the simulated astrophysical signals as noise triggers, damaging its sensitivity. The lack of triggers beyond $\chi_\mathrm{r}^2=10$ is due to our choice of vetoing such triggers.}
        \label{fig:chisq}
    \end{center}
\end{figure*}

\subsubsection{Single-detector signal-glitch discriminator}
\label{sec: signal-glitch}

In the presence of wide-sense stationary Gaussian noise, a consistently high \ac{SNR} across the detector network would have sufficed to assess the presence of \acl{GW} in the data. Advanced LIGO noise is, however, known to be neither wide-sense stationary nor Gaussian~\citep{Cabero:2019zyt}. Instead, the detector data contains short-duration noise transients or glitches that can produce large \acp{SNR}, mimicking transient \acl{GW}, therefore significantly damaging the search sensitivity. Consequently, it is necessary to implement signal-glitch discrimination techniques or ``vetoes'' to identify and penalize such glitches. PyCBC-based searches check whether a trigger: (a) is an outlier in the calibrated whitened data stream~\citep{Usman:2015kfa}, (b) has a morphology that is consistent with the best-matched template~\cite{Allen:2004gu}, (c) has any excess power beyond the maximum frequency of the best-matched template ~\cite{Nitz:2017lco}, (d) has any excess power (summed over bands) on particular timescales~\cite{Mozzon:2020gwa} and (e) is consistent across detectors~\cite{Usman:2015kfa, Nitz:2017svb, Davies:2020tsx}. 

These tests' outputs are numbers used to appropriately amend the trigger \ac{SNR} to suppress the noisy triggers. For instance, the output of the $\chi_\mathrm{r}^2$ test (b above) is commonly combined with the SNR to yield the ``re-weighted \ac{SNR}'':
\begin{equation}
    \tilde{\rho} =
    \begin{cases}
        \rho \Big[\frac{1}{2} \big(1+(\chi^2_\mathrm{r})^{3}\big)\Big]^{-1/6} & \mathrm{for\,} \chi^2_\mathrm{r} > 1 \\
        \rho & \mathrm{for\,} \chi^2_\mathrm{r} \leq 1
\end{cases}
\label{eq:newsnr}
\end{equation}
This statistic down-ranks triggers whose morphology is not consistent with the template~\citep{Babak:2012zx}, characterized by values of the $\chi_\mathrm{r}^2 > 1$. Such a situation can occur if the trigger is due to noise or it is due to an astrophysical signal that is not well-modeled by the template, as will be the case when signals with higher-order modes are filtered with templates that do not contain these. 

In Figure~\ref{fig:chisq} we have plotted $\chi^2_\mathrm{r}$ values as a function of the \ac{SNR} for a set of simulated waveforms that have been added to a representative section of Advanced LIGO Livingston data. These simulated waveforms are generated using the \texttt{SEOBNRv4HM} model, corresponding to signals from simulated binaries within the respective target spaces of the two template banks. The parameters of these simulated sources are uniformly distributed in $M_T(1+z)$ and $m_1/(m_1+m_2)$ space, isotropically across the sky sphere, uniformly over the polarisation angles and uniformly in comoving volume between bounds of $0.26~\mathrm{Gpc}^3$ and $40~\mathrm{Gpc}^3$. The top panels correspond to waveforms within the space spanned by ${\cal{B}}_1$, while the bottom ones correspond to ${\cal{B}}_2$. The left panels show the results of our searches, while the right ones correspond to the case where we use the PyCBC-IMBH search. It is noticeable that for the same \ac{SNR} the latter search returns much larger values of the $\chi_\mathrm{r}^2$ than ours as the templates used by it are the dominant harmonics of a quasi-circular \acl{BBH} as against the former where higher harmonics are present in the template waveforms. This will lead to a ``false identification'' of signals as glitches, damaging the search sensitivity. For better separation between the background triggers and the simulation, we discard triggers that are highly inconsistent with the templates by placing a threshold of 10 on $\chi^2_\mathrm{r}$ value. Additionally, we put a similar threshold on the short-term variation of the noise \acl{PSD}~\citep{Mozzon:2020gwa} following \citet{Chandra:2021wbw} to alleviate the effects of loud broadband detector noise.

\subsubsection{Background-Dependent Reweighing}

\begin{figure}[htb]
    \begin{center}
    \includegraphics[width=\columnwidth]{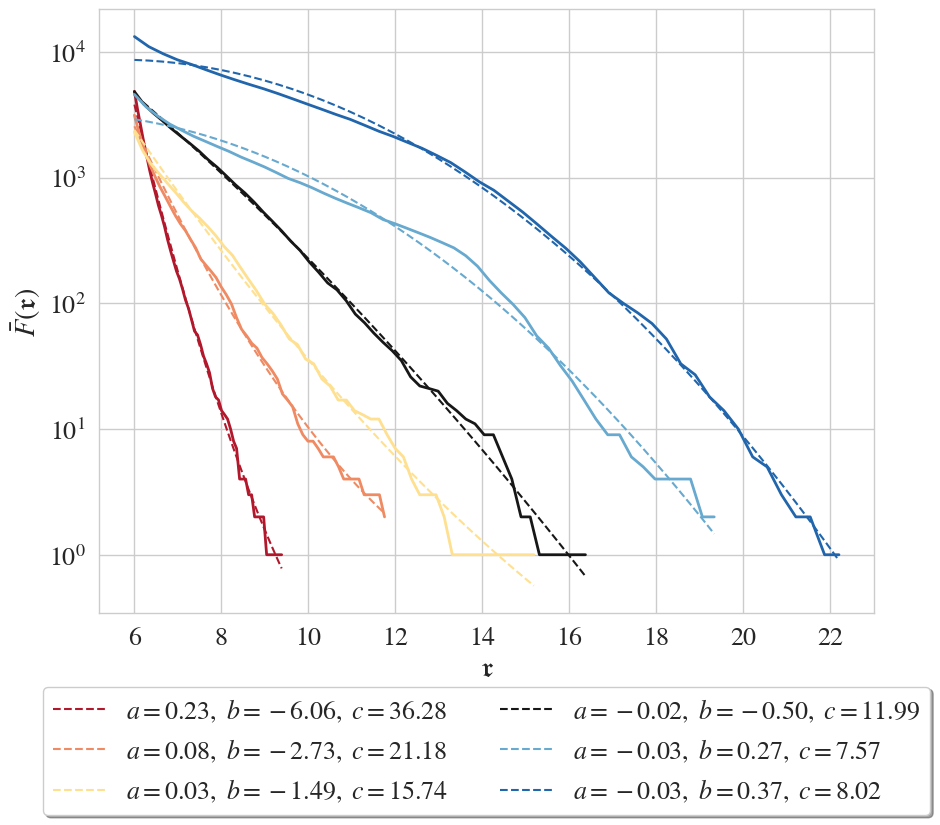}
    \caption{Complementary cumulative distribution of single detector triggers (solid lines)  binned by detector-frame total mass $M_T(1+z)$, effective inspiral spin parameter $\chi_\mathrm{eff}$ and symmetric mass ratio $\eta=q/(1+q)^2$, fitted with an exponential polynomial distribution (dotted lines) parameterized by $a,~b$ and $c$. The triggers correspond to those identified in a representative portion of Advanced LIGO Livingston data from the third observing run by a representational set of template groups.}
    \label{fig:fit}
    \end{center}
\end{figure}

The result of these consistency tests is a set of numbers, $v_i$ that are combined with the \ac{SNR} to produce a single-detector statistic $\mathfrak{r}(\rho;v_i)$~\footnote{For details see Sec.II C of~\citet{Chandra:2021wbw} and references therein.}. We expect the probability distribution of $\mathfrak{r}$ for triggers associated with noise in each detector $I$ beyond a threshold $\mathfrak{r}_0$ for each template $k$ is a falling exponential, with template-dependent parameter $a$. The complementary cumulative distribution of such a probability distribution is~\citep{Nitz:2017svb}
\begin{equation}\label{eq:exponential}
    \bar{F}_1(\mathfrak{r}) = e^{-a (\mathfrak{r}-\mathfrak{r}_0)}~.
\end{equation}
The above follows from the fact that the noise probability obeys an inhomogeneous Poisson process and that different templates identify different types of triggers. Consequently, certain regions of the bank will produce large values of $\mathfrak{r}$, potentially reducing the statistical significance of triggers coming from regions of the bank less prone to loud glitches. Therefore, we use an additional template-dependent parameter, $\mu_I^k$, that accounts for the total noise-trigger rate in a detector. The two are combined to obtain a model for the noise rate density for each template~\citep{Nitz:2017svb}:
\begin{equation}
    r_I^k = - \mu_I^k~\partial_\mathfrak{r}\bar{F}_1(\mathfrak{r}) = \mu_I^k a e^{-a (\mathfrak{r}-\mathfrak{r}_0)}
\end{equation}
which is used to separate the noise triggers from signal triggers further. Therefore template-dependent background reweighing uses the likelihood of a template to identify a trigger associated with noise of certain loudness to discriminate it. Generally, this fit could be performed separately for each template, but PyCBC-based searches choose to perform the fitting over a group of templates that behave similarly in the presence of noise. This increases the number of triggers for the fit and hence improves the fit.

However, we find that templates with higher harmonics produce noise trigger distribution that is not well reproduced by Eq.~\ref{eq:exponential}. Instead, we find experimentally that the noise probability is better fitted to a model whose complementary cumulative distribution is:
\begin{equation}\label{eq:ccdf}
    \Bar{F}_2(\mathfrak{r}) = e^{-[a (\mathfrak{r}-\mathfrak{r}_0)^2 + b (\mathfrak{r}-\mathfrak{r}_0) + c]}
\end{equation}
We show this resulting fit in Figure~\ref{fig:fit}. The solid lines correspond to single-detector triggers we obtained while performing the search in $\sim 30$ days of \ac{O3} data from Advanced LIGO-Livingston. The triggers are grouped based on the templates' total mass, effective spins, and symmetric mass ratio. The dotted lines are the estimated fits that follow the noise trigger distribution well. We note that the triggers correspond to a representative set of templates. For example, the blue line shows the distributions of triggers identified by templates with $M_T(1+z) \in (400, 500)M_\odot$, $\eta=q/(1+q)^2\in (0.11,~0.12)$ and $\chi_\mathrm{eff}\in(-0.6,-0.4)$ while the red line shows the same for templates with $M_T(1+z) \in (100, 200)M_\odot$, $\eta=q/(1+q)^2\in (0.08,~0.09)$ and $\chi_\mathrm{eff}\in(0.7,0.9)$.

We use these noise rate densities to estimate the total rate, $r_N^k$, of coincident events in a template. Under the assumption that the noise in both detectors is independent, this total noise rate estimate is proportional to the product of the rate of noise triggers in each detector for a template:
\begin{equation}\label{eq: network noise rate}
    r^k_N = - \prod_{I} \mu_{I}^k \partial_\mathfrak{r} \bar{F}_2(\mathfrak{r})
\end{equation}
One can use this total noise rate estimate at the multi-detector level to improve the statistical significance of astrophysical candidates, as described in the next subsection.

\subsubsection{Multi-Detector Ranking Statistics}

Under the assumption that the detector noise is Gaussian and uncorrelated, it is sufficient to use the network \acp{SNR}, $\sqrt{\rho_\mathrm{H}^2+\rho_\mathrm{L}^2}$, to rank coincident triggers from Hanford and Livingston detector~\citep{Pai:2000zt}.  Similarly, in the presence of non-gaussian glitches, the expression $\sqrt{\mathfrak{r}_\mathrm{H}^2+\mathfrak{r}_\mathrm{L}^2}$  provides a more suitable ranking of the coincident triggers ~\citep{Babak:2012zx}. We improve upon this by using a ranking statistic, $\mathcal{R}$~\citep{Nitz:2017svb}:
\begin{equation}
    \mathcal{R} \propto \frac{1}{\sqrt{2}}(\mathfrak{r}_{H,0} + \mathfrak{r}_{L,0} - \ln{r}_N^k)
\end{equation}
that includes the network noise rate estimate for a given template. It is designed to reduce to:
\begin{equation}
    \mathcal{R} \propto \frac{1}{\sqrt{2}}(\mathfrak{r}_H + \mathfrak{r}_L + \mathcal{O}(\mathfrak{r}^2))
\end{equation}
when the fit coefficient $b$ in both detectors is one. The pre-factor of $1/\sqrt{2}$ follows from the fact that if the rescaled \acp{SNR} in both the detectors is the same, then $\sqrt{\mathfrak{r}_H^2 + \mathfrak{r}_L^2} \sim \sqrt{2}\mathfrak{r}_H$. In both detectors, improvements to the ranking statistic using the relative probability distribution of nuisance parameters, such as $u$ and $A$, are left for future work.

\subsection{Estimation of statistical significance}

\begin{figure*}[thb]
    \centering
    \subfigure[Ratio of sensitive VT of our search-1 implementing the bank ${\cal{B}}_1$ for and the PyCBC-IMBH search.]{\label{fig:vt-1}
        \centering
        \includegraphics[width=0.9\textwidth]{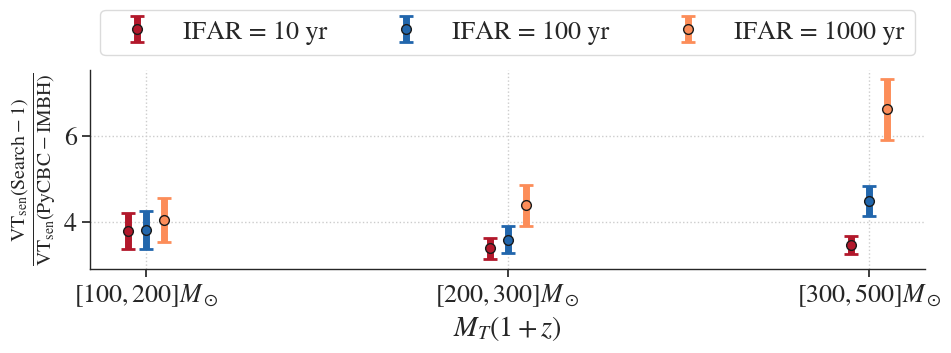}
    }
    \subfigure[Ratio of sensitive VT of our search-1 implementing the bank  ${\cal{B}}_2$ and the PyCBC-IMBH search.]{\label{fig:vt-2}
        \centering
        \includegraphics[width=0.9\textwidth]{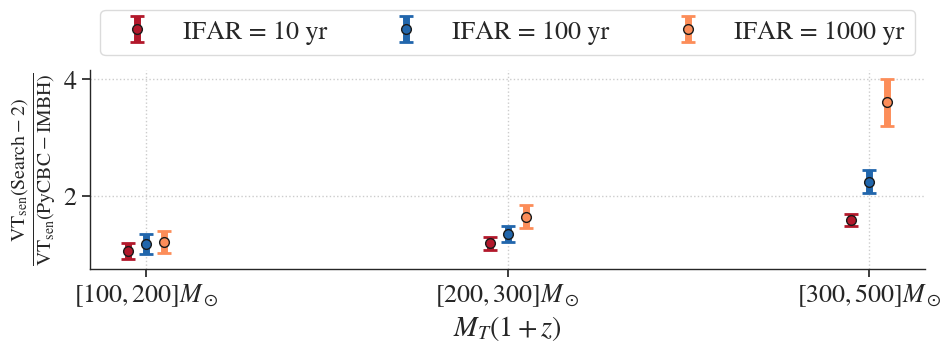}
    }
    \caption{Ratio of the volume-time sensitivity (VT) at a fixed inverse-false-alarm rate (IFAR) of the search presented in this work, which includes higher-order modes, with respect to the existing PyCBC-IMBH search, which omits higher modes. We show results as a function of the total redshifted mass of the target signals. The upper panel shows results corresponding to our low mass-ratio bank $\cal{B}_1$ targeting mass ratios $q \in (1,5)$ while the bottom one corresponds to our large mass-ratio bank $\cal{B}_2$ targeting mass ratios $q \in (5,10)$. In both cases, our search improves on existing ones, especially for short-lived high-mass target signals.}
    \hfill
\end{figure*}

We use $\mathcal{R}$ to estimate the statistical significance of coincident triggers. Ideally, we would like to calculate it using data free of \aclp{GW}. But there is no way to shield our detectors from incoming \aclp{GW}. However, it is possible to destroy the signal coherence across the detector network by shifting the data from one of the detectors with respect to other detector(s) by a time duration greater than the light travel time between the detectors~\citep{LIGOScientific:2016aoc}. Because the noise in each detector is assumed to be uncorrelated, time slides preserve the noise properties. It also helps simulate ``background triggers'', none of which are due to a given astrophysical signal. The distribution of the ranking statistic of these background triggers is used to compute \acfp{IFAR}, and any ``foreground'' or ``zero-lag'' trigger that exceeds a pre-determined \ac{IFAR} threshold is deemed as a detection candidate. 

Finally, since there can be chance coincidences between noise and signal triggers, our search hierarchically removes loud triggers like other PyCBC-based searches to minimize their impact~\citep{Usman:2015kfa} as well as background triggers that lie within a given time window around events with large \ac{IFAR}. 

\section{Search Sensitivity}
\label{sec:sensitivity}

To assess the benefits of our new search against searches that do not use higher-order mode waveforms, we estimate the sensitive volume-time product, $\mathrm{VT}_\mathrm{sen}$, using a set of simulated waveforms that we have added to $\sim 100$ days of O3 Advanced LIGO data. As described in Sec.~\ref{sec: signal-glitch}, our simulation set consists of binary black-hole waveforms with strong higher-order mode content. We compare our search performance against the PyCBC-IMBH search, used to search for intermediate-mass black-hole mergers in O3 data whose templates cover a parameter space including ours but ignore higher-order modes. We choose to compare the volumetric search sensitivity against PyCBC-IMBH~\citep{Chandra:2021wbw} as this search performs better than or comparable to other searches for intermediate-mass black hole binaries~\citep{Chandra:2021xvs}.

The sensitive volume-time product of a given search measures the number of expected signals from a population of binaries that the search can detect beyond a pre-determined statistical significance. We estimate this sensitive volume time using the Monte Carlo method with importance sampling. This involves sampling from a proxy or proposal distribution rather than the nominal or target distribution such that most of the simulated signals closely straddle the boundary between being detected and being missed. One can find details of the method used in the Appendix of \citet{Capano:2016dsf}.

 Figure~\ref{fig:vt-1} \& Figure~\ref{fig:vt-2} compare our searches' sensitive volume time estimate to the estimate from the PyCBC-IMBH search at different thresholds of \ac{IFAR}. The first plot is for a population of \aclp{BBH} with $q \lesssim 5$ while the latter is for binaries with $q \in (5,10)$, and we have used search-1 and search-2, respectively. In both cases, and independently of the \ac{IFAR} threshold, we observe that the searches implementing higher modes are more sensitive than those ignoring them. At low \ac{IFAR} a clear separation from injections and glitches is unnecessary, and the SNR recovery dominates the ranking statistic. Therefore, the difference in sensitivity is mainly driven by the difference in fitting factor caused by the lack of higher modes in the quadrupole search. In contrast, at large \ac{IFAR}, better separation between glitches and injections is needed. Therefore the detected simulated signals need to pass all the signal-glitch discriminator checks, e.g., the $\chi_\mathrm{r}^2$ test. For this reason, the quadrupole search further downweighs, resulting in an increased relative sensitivity of the higher-mode search.

The observed strength of the higher-order modes in the simulated signals grows as the total mass increases. Also, most of the in-band signal is dominated by the merger and ringdown stages. For this reason, we observe that the higher-mode search yields a larger sensitivity gain for increasing total mass. In particular, the sensitivity gain of search-1 (search-2) goes from a factor of $3.81\pm 0.44$ $(1.18\pm 0.17)$ for $M_T(1+z) \in (100,~200) M_\odot$ to $4.48 \pm 0.35$ $(2.25 \pm 0.20)$ for $M_T(1+z)\in[300,500]M_\odot$. We note that the reduced sensitivity gain of the search-2 is because its bank, ${\cal{B}}_2$, has a significantly larger amount of templates and, therefore, suffers from an increased background.

\begin{table*}[htb]
    \centering
    \begin{tabular}{llllllll}
         \multirow{2}{*}{Event Name} & \multirow{2}{*}{GPS Time~[s]}~& \multicolumn{3}{c}{Search-1} & \multicolumn{3}{c}{Search-2} \\
         & & $\mathrm{IFAR}~[\mathrm{yr}]$ & $\rho_\mathrm{H}$ & $\rho_\mathrm{L}$ & $\mathrm{IFAR}~[\mathrm{yr}]$ & $\rho_\mathrm{H}$ & $\rho_\mathrm{L}$\\
         \toprule
          $\GW190408\_181802$ & $1238782700.3$ & 3.44 & 7.74 & 7.80 & 60.60 & 9.04 & 8.02 \\
          $\GW190412$ & $1239082262.2$ & - & - & - & 10.50 & 6.81 & 10.63 \\
         $\GW190503\_185404$ & $1240944862.3$ & 3.52 & 8.81 & 7.41 & - & - & -\\
         $\GW190513\_205428$ & $1241816086.8$ & 1.49 & 7.31 & 6.95 & - & - & -\\
         $\GW190517\_055101$ & $1242107479.8$ & 1.88 & 5.97 & 7.82 & - & - & -\\
         $\GW190519\_153544$ & $1242315362.4$ & 31.35 & 8.52 & 9.56 & 18.82 & 7.07 & 8.96\\
         $\GW190521$ & $1242442967.4$ & 1.13 &7.56 & 11.85 & - & - & -\\
         $\GW190521\_074359$ & $1242459857.4$ & 43069.81 & 10.52 & 19.18 & 598.19 & 8.97 & 17.05 \\
         $\GW190602\_175927$ & $1243533585.1$ & 2.07 & 6.49 & 10.16 & - & - & -\\
         $\GW190706\_222641$ & $1246487219.3$ & 108.26 & 8.80 & 8.24 & - & - & -\\
         $\GW190727\_060333$ & $1248242632.0$ & 27.65 & 7.50 & 6.95 & - & - & -\\
         $\GW190828\_063405$ & $1251009263.8$ & 12390.23 & 8.50 & 9.64 & 210.00 & 9.11 & 10.23 \\
         $\GW190915\_235702$ & $1252627040.7$ & 16.83 & 7.49 & 6.88 & 9.80 & 8.87 & 9.66 \\
         $\GW191109\_010717$ & $1257296855.2$ & 5.43 & 8.52 & 12.25 & - & - & -\\
         $\GW200128\_022011$ & $1264213229.9$ & 2.44 & 6.41 & 6.32 & - & - & - \\
         $\GW200224\_222234$ & $1266618172.4$ & 1769.27 & 10.59 & 10.77 & 106.05 & 10.90 & 10.56 \\
         $\GW200225\_060421$ & $1266645879.4$ & 6.08 & 7.63 & 6.09 & 59.95 & 7.85 & 6.38 \\
         $\GW200311\_115853$ & $1267963151.3$ & 2501.38 & 9.70 & 8.94 & 533.38 & 9.85 & 9.25\\
    \hline
    \end{tabular}
    \caption{\acl{GW} events recovered by our search with \acl{IFAR} $>1$ year, calculated using \ac{O3} data from Hanford and Livingston detector. These events have all been reported in~\citet{LIGOScientific:2020ibl, LIGOScientific:2021djp, Nitz:2021uxj, Nitz:2021zwj, Chandra:2021wbw, Olsen:2022pin} with a comparatively higher statistical significance. We attribute their comparatively lower significance to their lack of detectable higher-order modes and/or being outside the bounds of our target search space. }
    \label{tab:events}
\end{table*}

\section{ Results from the third LIGO observing run}
\label{sec:results}

Given the sensitivity improvement to massive \acl{BBH} signals with higher harmonics, we deployed our new search to analyze the publicly available \ac{O3} data from Advanced LIGO detectors~\citep{LIGOScientific:2019lzm}. To do this, we divided the data into nine independent blocks of $\sim 30$-day duration and analyzed it using our template banks. We report all the \ac{O3} candidates reported by our searches with an \acl{IFAR} $>1$ year in Table~\ref{tab:events}, finding no new candidates beyond those reported in ~\cite{LIGOScientific:2020ibl, LIGOScientific:2021djp, Nitz:2021uxj, Nitz:2021zwj, Olsen:2022pin}. We note that the (expected) reduced statistical significance of the reported events with respect to those obtained by existing searches highlights that these candidates are outside our target space.

We note that the lack of new detections is not unexpected. Firstly, signals from binaries with the inclinations targeted by our banks are significantly weaker than those from face-on binaries typically detected by existing searches. Secondly, in contrast to existing searches, we have performed a two-detector search using only Advanced LIGO Hanford and Advanced LIGO Livingston data, reducing the loudness of the signals across the detector network. We have checked that these two effects lead to an optimally observable volume  $\simeq 7$ times smaller than that of search targeting rather than face-on signals. The fact that existing searches have only reported three events to date within our targeted mass range (GW190521, GW200220, and GW190426) \cite{LIGOScientific:2021djp} is perfectly consistent with our results. In addition, the inclusion of a third -- Virgo -- detector in our search will also improve our ability to discriminate glitches from true signals, further increasing our sensitivity. We leave the inclusion of further detectors in our search for further improvement. Such work will primarily involve extending the two-detector ranking statistic to a multi-detector one.

\section{Conclusions}
\label{sec:conclusions}

We have presented the first matched-filter gravitational-wave search for compact mergers, including the impact of higher gravitational-wave modes. On the one hand, detecting merger-ringdown higher modes is crucial to enabling tests of General Relativity in the strong-field regime and is key to observing several strong-field phenomena arising during the merger-ringdown stages. On the other hand, due to the high mass of these systems, the emission from intermediate-mass black-hole mergers contains strong higher-mode contributions for orientations other than face-on/off. This makes this type of search crucial for both fundamental physics and astrophysics.

While one can extend our search to signals from any binary merger, we have specifically restricted to those with strong higher-mode content: asymmetric, non-precessing, massive black-hole binaries at large orbital inclinations, leaving the extension to more generic inclinations as future work. While we find no new candidates with \ac{IFAR} $>1$ year beyond those already reported elsewhere, we have demonstrated that our search is up to 450\% more sensitive than existing matched-filter searches with overlapping parameters space but omitting the higher-mode content. We stress that the lack of new, higher-mode rich detections is somewhat expected, as the highly inclined systems we target approximately span an observable volume $\simeq 7$ times smaller than the rather face-on ones targeted by existing searches, which have detected three events in our target mass-range to date.

Also, the expected merger rate density of \acl{IMBH} binaries is $< 1/\mathrm{Gpc}^3/\mathrm{yr}$ as compared to $\sim 23/\mathrm{Gpc}^3/\mathrm{yr}$ for stellar-mass black hole binaries~\citep{LIGOScientific:2021tfm, LIGOScientific:2020kqk}. This means that we expected to observe $\sim 100$ stellar-mass black holes as compared to $\sim 3$ \acl{IMBH} binaries during O3 given that the achieved sensitivity allowed us to observe $30M_\odot+30M_\odot$ and $100M_\odot+100M_\odot$ to a distance of $\sim 1$Gpc and $\sim 2$Gpc respectively. As the detector's distance reach improves, we are likely to observe more \acl{IMBH} binaries, and thus, we are more likely to observe signals from our target population in the future~\citep{KAGRA:2013rdx}.

Finally, we note that our existing search needs to undergo several improvements before being equivalently mature with respect to existing ones. First, our search is restricted to data from only two detectors. Extending this to an arbitrary number of detectors would greatly help better discriminate true signals from glitches. Second, there is room to improve how we exploit the information contained in higher harmonics to remove more background triggers. Third, our search at the moment is meant to complement existing searches with overlapping target parameter space. A near-term development goal is to devise a strategy that combines the search output of the contributing searches so that we quote a single quantitative estimate of an event's statistical significance. Last, as it is also the case for other existing searches, an obvious extension would be expanding the target search space to include orbital precession or eccentricity. The first is straightforward to implement, given the genericness of our search method, while the others require a combination of stronger vetoes~\citep{Dhurandhar:2017aan, Jadhav:2020oyt, McIsaac:2022odb, Davis:2022cmw}, glitch-robust search statistics and glitch subtraction~\cite{Cornish:2020dwh}, that can exploit the complex waveform morphology due to addition of new physics.

\section*{Acknowledgements}
\label{sec:acknowledgements}
%
We thank Tito Dal Canton, Thomas Dent, and Kritti Sharma for their detailed comments and valuable suggestions. We would also like to thank Ver\'{o}nica Villa-Ortega for running an earlier version of the presented algorithm. This material is based upon work supported by NSF's LIGO Laboratory, a major facility fully funded by the National Science Foundation. This research has used data or software from the Gravitational Wave Open Science Center (gw-openscience.org), a service of LIGO Laboratory, the LIGO Scientific Collaboration, the Virgo Collaboration, and KAGRA. LIGO Laboratory and Advanced LIGO are funded by the United States NSF as well as the STFC of the United Kingdom, the Max-Planck-Society (MPS), and the State of Niedersachsen/Germany for support of the construction of Advanced LIGO and construction and operation of the GEO600 detector. The Australian Research Council provided additional support for Advanced LIGO. The authors are grateful for the computational resources and data provided by the LIGO Laboratory and supported by National Science Foundation Grants No. PHY-0757058 and No. PHY-0823459. The authors also acknowledge using the IUCAA LDG cluster, Sarathi, for computational/numerical work. KC acknowledges the MHRD, the Government of India, for the fellowship support. JCB is supported by a fellowship from the ``la Caixa'' Foundation (ID100010434) and from the European Union’s Horizon 2020 research and innovation program under the Marie Skłodowska-Curie grant agreement No 847648. The fellowship code is LCF/BQ/PI20/11760016. JCB is also supported by the research grant PID2020-118635GB-I00 from the Spain-Ministerio de Ciencia e Innovaci\'{o}n. AP's research is supported by SERB-Power fellowship grant SPF/2021/000036, DST, India. IWH acknowledge the STFC for funding through grant ST/T000333/1 and ST/V005715/1. This document has LIGO DCC No LIGO-P2200182.

\end{document}